\documentclass[referee]{raa}            

\usepackage{graphicx,times}             
\usepackage{natbib}
\usepackage{amssymb,amsmath}
\bibpunct{(}{)}{;}{a}{}{,}

\usepackage[a4paper=true,dvipdfm=true,pagebackref=true]{hyperref}

\begin{document}

   \title{Fast Radio Bursts as crustal dynamical events induced by magnetic field evolution in young magnetars
}

   \volnopage{Vol.0 (20xx) No.0, 000--000}      
   \setcounter{page}{1}          

   \author{J.E. Horvath
      \inst{1}
   \and P.H.R.S. Moraes
      \inst{1}
      \and M.G.B. de Avellar
      \inst{2}
   \and L.S. Rocha
      \inst{1}
   }

   \institute{Universidade de S\~ao Paulo (USP), Instituto de Astronomia, Geof\'isica e Ci\^encias Atmosf\'ericas, Rua do Mat\~ao 1226, Cidade Universit\'aria,
05508-090 S\~ao Paulo, SP, Brazil; \,\, {\it foton@iag.usp.br}\\
        \and
             Universidade Federal de S\~ao Paulo (UNIFESP),
Instituto de Ci\^encias Ambientais, Qu\'imicas e Farmac\^euticas,
Rua S\~ao Nicolau 210, 09913-030 Diadema, SP, Brazil}

\vs\no
   {\small Received~~20xx month day; accepted~~20xx~~month day}

\abstract{ We revisit in this work a model for repeating Fast Radio Bursts based of the release of energy provoked by
the magnetic field dynamics affecting a magnetar's crust. We address the basics of such a model by solving the propagation of the perturbation approximately, and quantify the energetics and the radiation by bunches of charges in the so-called {\it charge starved} region in the magnetosphere. The (almost) simultaneous emission of newly detected X-rays from SGR 1935+2154 is tentatively associated to a reconnection behind the propagation. The strength of $f$-mode gravitational radiation excited by the event is quantified, and more detailed studies of the non-linear (spiky) soliton solutions suggested.
\keywords{Fast Radio Bursts --- Magnetars --- Alfven waves}
}

   \authorrunning{J.E. Horvath et al. }            
   \titlerunning{FRBs from magnetic quakes in young magnetars }  

   \maketitle

%
%
\section{Introduction}           
\label{sect:intro}

The last decade has witnessed the emergence of the study of a class of transients which keep the community very active in search for answers.As a relatively recently observed phenomenon, Fast Radio Bursts (FRBs) are bright pulses emission at radio frequencies which last for $\sim$ms or even less (Petroff et al. 2019, Katz 2018, Popov, Postnov \& Pshirkov 2018). Their emission has been detected in the interval $400$MHz-$8$GHz, considered as ``typical'', with at least one case in which slightly delayed X-ray bursts coincide with the radio spikes (Mereghetti et al. 2020).

The first ever detected FRBaccepted as so, later named as FRB 010724, was discovered by Lorimer et al. (2007) in surveys of radio pulsars (Beskin 2018). For years, such a discovery remained the only known signal of its kind. Strong support to a short-duration transient with similar characteristics to the burst reported in Lorimer et al. (2007) became available in 2013 only, when Thornton et al. (2013) reported the observation of four high-dispersion measure pulses with the Parkes Telescope facility (Keith et al. 2010). 	

Over the years, more and more FRBs have been detected, not only by Parkes Telescope, but also by Arecibo Observatory (Spitler et al. 2014), Australian Square Kilometer Array Pathfinder (ASKAP) (Bannister et al. 2017), Canadian Hydrogen Intensity Mapping Experiment (CHIME) (CHIME/FRB Collaboration 2019a) and others. The high values of the dispersion measure of FRBs mentioned above strongly indicate that they have extragalactic or cosmological origin, a conjecture which is indeed confirmed by most FRB observations. It became clear later that FRBs are quite luminous outbursts, with luminosity $L\sim10^{43} \,erg/s$ if arising from extragalactic sources, and much consideration and interest have been given to them, among other reasons due to the amazing mechanism behind the progenitor systems of such extreme events that must be operating.

In 2016 the first {\it repeating} FRB was observed, the FRB 121102 (Spitler et al. 2016). Possibly, repeating FRBs come from an entirely different source class or progenitor system compared to the non-repeating population, if the latter indeed exists. It is not known whether {\it all} sources repeat, because the repeating times may be long, thus the need of some so-called ``cataclysmic'' models for the non-repeating sources is not yet compelling. Some models able to explain the repeating FRBs, such as FRB 121102 and also the later reported FRB 180814.J0422+73 (CHIME/FRB Collaboration 2019b), are: relativistic beams accelerated by impulsive magnetohydrodynamic driven mechanism, which interact with clouds at the center of star-forming dwarf galaxies (Vieyro et al. 2017); soft-gamma repeaters (Wang \& Yu 2017);  starquakes of a magnetar (Wang et al. 2018); mass transfer in magnetic white dwarf and NS with strong bipolar magnetic field binary systems (Gu et al. 2016), highly magnetized pulsars traveling through asteroid belts (Dai et al. 2016) and binary neutron stars not far away from merging (Totani 2013, Wang et al. 2016). Several other proposals can be seen, for instance, in Wadiasingh et al. (2020), Dai et al. (2016), Wadiasigh and Timokhin (2019), Gupta and Saini (2018), Michilli et al. (2018), Levin, Belobodorov and Bransgrove (2020), Ioka and Zhang (2020), Kashiyama and Murase (2017), a state that heavily reminds the situation of GRBs in the decade before 1990s (Nemiroff 1994) and suggests that more than one event is actually contributing to the FRB phenomenon. The issue of repeating vs. non-repeating sources is actively under discussion. Given the recent detections and localizations of the non-repeating FRB 190824 (Bannister et al. 2019) and FRB 190523 (Ravi et al. 2019), Gourdji et al. (2020) have placed constraints on the magnetic field strength of the putative-emitting NS. In order to explain the FRB 150418 (Keane et al. 2016), Zhang has invoked a merging NS-NS binary producing an undetected short gamma-ray burst and a supra-massive NS, which subsequently collapses into a black hole (Zhang 2016b). Binary black holes were also considered, in spite of the general thought that electromagnetic signals should not emerge from these events (Zhang 2016a, Liu et al. 2016).

If FRBs are indeed related to NSs and/or black holes, it is natural to consider them in current and future gravitational wave searches. Callister, Kanner and Weinstein (2016) have shown that Advanced LIGO (Laser Interferometer Gravitational wave Observatory) and Virgo observations can severely constrain the validity of FRB binary coalescence models. FRBs can also provide an unprecedented tool for observational cosmology. In Caleb, Flynn and Stappers (2019), for instance, FRBs were used for tracing the \texttt{He II} reionization epoch from simulations of their dispersion measures. Finally, in Wei, Wu and Gao (2018) it was proposed that upgraded standard sirens can be constructed from the joint measurements of luminosity distances derived from gravitational waves and dispersion measures derived from FRBs. Such an upgrade has shown to be more effective for constraining cosmological parameters. These considerations are especially important for non-repeating, catastrophic events models of FRBs.

On the other hand, an important report of repeating FRBs associated with a galactic magnetar source (Bochenek et al. 2020) SGR 1935+2154 has confirmed that NSs with high magnetic fields are involved in the production of FRBs. The magnetar has been associated with the supernova remnant SNR G57.2+0.8 with age in excess of $\sim 10^{4} yr$ (Zhou et al. 2020), although this may be misleading if the object inserted energy into the remnant at birth (Allen \& Horvath 2004), because this makes the remnants look much older that they really are. However, the recent work of (Zhou et al. 2020)  did not find evidence in favor of a large energy injection. Independently of this last consideration, it is clear that a successful model for the production of FRBs in magnetars is needed.

We address here a model in which FRBs are originated from NS crustal events induced by magnetic field evolution in young magnetars. As a supporting fact, it is worth to mention that in Wang et al. (2018), the burst rates of FRB 121102 were shown to be consistent with a kind of seismicity rates. However, the scenario proposed in Wang et al. (2018) consists of a magnetar with a solid crust in which the stellar shape changes from oblate to spherical, what induces  stresses in the crust, yielding a starquake.

We argue below that it is the behavior of the magnetic field, suffering from instabilities in the first few centuries after birth that may produce i) the free-energy source ; ii) the right timescales and iii) the perturbations at the crust's plates that suitably shake the field lines, sending Alfven waves/solitons into the magnetosphere, resulting in the production of curvature radiation that may be the origin of the detected FRBs with the required frequency. The production of associated X-ray bursts and the possibility of detection of $f$-modes from the crust induced by these events are also briefly addressed.

\section{Magnetic field evolution in young magnetars}\label{sec:frbq}

The group of magnetar NSs, in which activity is supposed to arise from the behavior of magnetic fields, includes the Soft-Gamma Repeaters and the Anomalous X-Ray Pulsars. A very high value of the magnetic field $B$ has been inferred for them, in spite that in a few cases (notably AXPs...) a much smaller field seems to be present (see Turolla \& Esposito 2013 and references therein). Since a pure poloidal field is known to be unstable, it is suspected that a toroidal component acting as a reservoir of free-energy is also present.

A relevant detailed investigation for this picture has been performed by Gourgouliatos, Wood and Hollerbach (2016). They have shown that in a plausible scenario of energy equipartition between global-scale poloidal and toroidal magnetic field components, magnetic instabilities transfer energy to non-axisymmetric, sub-$km$-sized features, in which local field strength can greatly exceed that of the global-scale field. Such intense small-scale magnetic features were shown to induce high-energy bursts through local crust yielding. Essentially, it was shown that the observed diversity in magnetar behavior can well be explained with mixed poloidal-toroidal fields of comparable energies.

Neutron starquakes have been considered as a model for gamma ray bursts long ago (Pacini \& Ruderman 1974, Epstein 1988, Blaes et al. 1989). In such models, elastic energy is released in a crustquake, exciting oscillations of the surface magnetic field. The induced electric field accelerates high energy particles, which in turn radiate gamma rays. The later discovery of the extragalactic character of the sources diminished the interest in quake models, now revived by the identification of SGR 1935+2154. Instead of the release of elastic energy, our scenario suggests the magnetic field as the {\it cause} of crust breaking, and the propagation of Alfven waves (ordinary or solitonic) along the field lines radiating the observed radio photons.

We start by the sort of basic energetic calculation that motivates the whole picture. Consider a solid NS crust in which a magnetic field of strength $B$ evolves on sub-$km$-sized patches with length $l$. The magnetic energy is

\begin{equation}
    E_{B} = 4 \times 10^{40} \, B_{15}^{2} \, l_{4}^{3} \,\,  erg,
\end{equation}
where we have scaled the quantities as $B_{15} \equiv (B/10^{15} G)$ and $l_{4} \equiv (l/10^{4} \, cm)$ (Gourgouliatos \& Esposito 2018). We note that giant magnetar flares have been modeled within a similar picture, although we believe that the latter may be
a different phenomenon (no giant flare has been associated with a FRB), perhaps a magnetospherically-originated instability.

On the other hand, the crust magnetoelastic energy is

\begin{equation}
E_{me} = 4 \times 10^{38} \, B_{15}^{-2} \, l_{4}^{2} \, \sigma_{-3}^{2} \,\, erg
\end{equation}
with $\sigma_{-3} \equiv (\sigma/10^{-3})$ being the dimensionless stress modulus (Thompson \& Duncan 1995). Detailed calculations have shown (Horowitz \& Kadau 2009, Hoffman \& Heyl 2012) that the crust cracks at a critical value of
$\sigma_{max} = 0.1$, so that from eqs.(1) and (2) we see that the critical field enough to crack the crust is

\begin{equation}
    B \sim 3 \times 10^{15} {\biggl( {\sigma \over{0.1}} \biggr)}^{1/2}.
\end{equation}

Since the local field exceeds the global value by an order of magnitude (Gourgouliatos, Wood \& Hollerbach 2016), it is the magnetic field that drives the energetics. The instabilities will induce a ``lifting'' of the cracked crust material of size $l$, typically displaced a distance $\Delta l \simeq 100 \sigma_{-3} \, l_{4} cm$.
A more accurate calculation (Lander et al. 2015) has employed a cracking condition based on the {\it local} von Mises criterion, i.e.,

\begin{equation}
    {\biggl( {1\over{2}} \sigma_{ij} \sigma^{ij} \biggr)}^{1/2} \, \geq \sigma_{max},
\end{equation}
and also attempted to predict where exactly the crust should crack. We refer to the work of Lander et al. (2015) and keep the simplest estimates in the following.

\section{Transmission of energy into the magnetosphere: linear and non-linear regime solutions}

The Alfven equation derived in the Appendix (eq.A.10)  has been studied before and, at least in its simplest form, admits the propagation of solutions for the perturbation $\xi$ of relevance for the FRB problem.
A brief description of this kind of propagating mode in in order. First of all, the amplitude of the Alfven wave is estimated to be $B/B_{15} \sim 10^{-4}$, therefore a linearization
seems justified. The physical picture is that an electric current is carried along the field lines by electrons and positrons moving in opposite directions. Bunches of particles with
approximately zero electric charge are formed, and along the propagation the velocity must speed up to compensate the decreasing density, as required to comply with the current density needed to sustain the wave. When the plasma density in the magnetosphere falls below a critical value, the propagation can no longer be supported, and a large electric field develops to compensate {\it via} the displacement current the decrease of the current density. This electric field boosts electrons and positrons to high Lorentz factors in opposite directions along the field lines. This
is when the synchro-curvature radiations of the bunches, ``summed up'' in the quantity $\xi$ is produced, as discussed, for example, by Kumar, Lu \& Bhatacchraya (2017). Therefore we
must solve the propagation of $\xi$ to calculate the emerging radiation.

We shall not treat the behavior in the crust because in our picture of magnetoquakes the instability of the field would provoke a lifting of a crust plate with a speed $v_{p}$ in which the field lines are frozen. Therefore, we shall focus on the magnetosphere propagation $z > 0$. Recalling the assumptions made when we wrote down eq.(A.10), in the region $z>0$ evidently $\mu = 0$, and we can also neglect the density $\rho$ compared to the $B^{2}$ term (by the very nature of a magnetosphere (Blaes et al. 1989).
We shall further assume $\cos \alpha$ constant, independent of $z$ to allow an analytical solution (we have checked that an expected dependence with $z$ of a dipolar $B$ field does not change much the picture. The latter case can be solved and the solutions happen to be linear combinations of the first-order Bessel functions $H_{1}$ and $Y_{1}$, with a slightly decreasing amplitude for large $z$.). The equation to be solved in this simplified case is just

\begin{equation}\label{11.1}
	\frac{d^{2}\xi}{dz^{2}} - \frac{d^{2}\xi}{{v_{A}^{2}}dt^{2}} = 0 \,\,,
\end{equation}
with $v_{A}^{2} = (c \, \cos\alpha)^{2}$. Because of the impulsive physical picture, we proceed to calculate the Green function $G(z,t)$ for this problem as a relevant step to understand the propagation of $\xi$. Assuming an inhomogeneous unitary perturbation $\delta(z) \delta(t)$, the solution must satisfy the initial conditions

\begin{eqnarray}
    \xi(z, t=0) = \xi_{0}(z), \label{}\\
     {\frac{\partial \xi}{\partial t}} (z,t=0) = v_{p}(z) \,\, \label{} .
\end{eqnarray}

Once the Green function (having two contributions $G^{0}$ and $G^{1}$) are found, the solution for $\xi$ reads

\begin{eqnarray}
\xi (z,t) = {\int G^{0}(z-z', t) \xi_{0}(z) dz'} + \nonumber\\
+ {\int G^{1}(z-z', t) {v_{p}}(z) dz'} \,\, \label{} .
\end{eqnarray}

Imposing the appropriate initial conditions and transforming to the Fourier space $G(z,t) = \int g(k,t) e^{i k.z} \frac{dz}{2\pi}$ the transformed amplitude $g(k,t)$ is the solution of the equation $\ddot g + k^{2}g =0$, that is
$g(k,t) = A(z) \cos kt + B(z) \sin kt$. Enforcing the two sets of initial conditions

\begin{eqnarray}
g^{0} (k,0) = 1   \,\,\,\,\, {\dot g^{0}} (k,0) = 0 \\
g^{1} (k,0) = 0   \,\,\,\,\, {\dot g^{1}} (k,0) = 1 \,\, \label{} ,
\end{eqnarray}
we obtain $g^{1} = \frac{1}{k} \sin{kt}$ and  $g^{0} = \cos{kt} \equiv {\dot g^{1}}$. Therefore the Green function is just

\begin{equation}
    G^{1}(z,t) = \int_{-\infty}^{\infty}  \frac{\cos{kz} \sin{kt}}{k}  \frac{dk}{2\pi} \,\, \label{} ,
\end{equation}
which is easily integrated using the symmetry $k \rightarrow -k$ to yield

\begin{equation}
    G^{1} (z,t) = {\frac{1}{4}} {\biggl[ {sign (v_{A}t + z) + sign (v_{A}t - z)} \biggr]} \,\, \label{} ,
\end{equation}
with $sign {x}$ being the {\it signum} function of the argument.
Now the convolution of eq.(8) with the perturbation having $\xi_{0} = 0$ makes unnecessary the $G^{0}(z,t)$, and we are left with the propagating
solution

\begin{eqnarray}
\xi(z,t) =\frac{1}{4} \biggl[ \int {sign(v_{A}t+(z - z'))v_{p}(z')dz'}\nonumber \\
+\int{sign(v_{A}t-(z - z'))v_{p}(z')dz'} \biggr]\,\, .
\end{eqnarray}

This is a very simple solution for the perturbation $\xi$ traveling along the field lines, but a particularly suitable one because a sudden, unitary perturbation is physically invoked as the origin of the propagating $\xi$ as stated above. The charges traveling {\it inside} $\xi$ could, in principle, radiate photons by synchrotron and curvature mechanisms as explained above. The general case, dubbed {\it synchro-curvature} radiation embedding both effects is, however, reduced to the curvature contribution in our case because high magnetic fields are known to cool very efficiently, reducing the radiative and conductive opacities $\kappa_{rad}$ and $\kappa_{cond}$ (Istomin \& Sobyanin 2007). For a single electron the curvature radiation is (Xiao, Wang \& Dai 2021)

\begin{equation}
    {dE \over {d {\omega} dt}}{\bigg|_{single}} = {{{\sqrt {3}} e^{2} \gamma} \over {2 \pi R_{c}}}
    {\omega \over {\omega_{c}}} {\int_{\omega/\omega_{c}}^{\infty} K_{5/3} (\mu) d \mu} \,\, \label{} ,
\end{equation}

where $R_{c}$ is the curvature radius, $\omega_{c} (3/2) (\gamma^{3} c /R_{c})$ a characteristic curvature frequency and $K_{5/3}$ is the modified Bessel function of second kind. To contribution to the total radiated power of the charges in the perturbation is essentially the coherent contribution of the charges travelling inside the perturbation $\xi$.

The radiation spectrum can be evaluated using eq.(14) and the solution eq.(13). On very general grounds, it is quite possible to understand the main features. It is well-known that the Bessel function $K_{5/3}$ admits an integral representation $K_{\alpha} (\mu) = \int_{0}^{\infty} e^{(- \mu \cosh t)} \cosh \alpha t  \, dt$, and this form immediately shows that the main contribution is given for low values of the dummy argument $\mu \approx 1$. In fact it is enough to use for the evaluation the approximate form $K_{\alpha} \rightarrow (\Gamma (\alpha)/2) {(2 / \mu)^{\alpha}}$ valid for $0 \leq | \mu | \leq \sqrt {(\alpha + 1)}$. The result is

\begin{equation}
    {dE \over {d {\omega} dt}} \approx  {{3{\sqrt {3}} e^{2} N^{2} \gamma} \over {2^{2/3} \pi R_{c}}}
    {\biggl( {\omega_{c} \over {\omega}} \biggr)}^{2/3} \,\, \label{} ,
\end{equation}

where the coherence signature, $N^{2}$ is the result of evaluating the so-called {\it phase stacking} integral at the relevant frequencies (Xiao, Wang \& Dai 2021). The size of the region in which coherence can be maintained scales as $1/\gamma$, and the condition of {\it charge starvation} has been generally considered as important for the radiation (Blaes et al. 1989, Chen et al. 2020). A more detailed treatment shows additional spectral features not treated in this approximate expression (Yang \& Zhang 2018).

An application of this formula to our problem yields a curvature radiation power below $\omega_{c}$ of

\begin{equation}
    {d E \over {dt}} = 2 \times 10^{38} {\biggl( {\gamma \over {100}} \biggr)} {\biggl( {n \over {10^{3} n_{GJ}}} \biggr)}^{2} erg \, s^{-1} \,\, ,
\end{equation}

where $n_{GJ} = 7 \times 10^{11} (B/10^{12} G) (P/0.1 s)^{-1} cm^{-3}$ is the Godreich-Julian charge density (Goldreich \& Julian 1969,  Rajagopal \& Romani 2007), chosen as a reference value. We see that unless the charge density inside the perturbation is substantially higher than the Goldreich-Julian value, there would be not enough power in the curvature radiation to match the observations of FRBs (Xiao, Wang \& Dai 2021). This high density may well be the case (Istomin \& Sobyanin 2007, Kunzl et al. 1998), but the present status of the knowledge of magnetospheric configurations of magnetars does not allow a firmer statement. However, it is important to note that an evaluation of the power-loss likely stays below the maximum energy released eq.(2), but could be enough to be in a strongly beamed pulse at the $\sim GHz$ frequency range, which could be associated to the FRB pulse.

We must point out that the general propagation of waves in the magnetosphere is not restricted to the Alfven linear waves of the type just  discussed.  The  ultra-relativistic  plasma  actually supports a variety of propagating modes which have to be studied case-by-case (Treumann \& Baumjohann 1997), without the linearization of the equations. We have not performed such a study here, however, we suggest that a class of these non-linear solutions, dubbed {\it spike solitons} may be of particular interest in this problem. Their rapid phase rotation has been associated to concrete observations in the solar wind, and they have been shown to exist in ultra-relativistic plasmas, of which the magnetospheres constitute a prime example.

Two important conditions of these solutions is that their amplitude is large at infinity, and their velocity approaches the light value $v_{A} \rightarrow c$. Spiky solitons have been pictured as sharp density ``humps'' with a length $\Lambda$ which travel along the field lines, which  produce curvature and additional radiation (called ``transition radiation'' in Ginzburg and Tsytovich (1979) when electrons or positrons from the ambient enter the hump region, a feature that is shared by the linear solutions discussed above. If we denote with $v$ the relative velocity of $N_{e}$ electrons or positrons that contribute to the transition radiation, its total power has been estimated as (Sakai \& Kawata 1980)

\begin{equation}
{dE\over{dt}} = {v\over{L}} N_{e}^{2} {e^{2} \omega_{p}\over{24 c}} {\epsilon_{0}\over{mc^{2}}}
{\biggl({{<B>}^{2} \over{16 \pi c^{2} (1-{v_{A}^{2} \over{c^{2}}})}}\biggr)}^{2} \,\, \label{} ,
\end{equation}

with $\omega_{p}= {\sqrt {(4 \pi n_{e} e^{2}/m_{e})}}$ the plasma frequency and $\epsilon_{0}$ the kinetic energy of one electron/positron, and $<B>$ is the average field swept by the spiky soliton along its path. With typical parameters, this power is, however, negligible and it can not contribute to a more isotropic, diffuse emission (Resmi, Vink \& Ishwara-Chandra 2020)
observed in the repeating magnetar source.

These estimates are quite crude, and a complete, explicit solution of the spiky solitons would be needed to properly evaluate the emission (for example, the actual length $\Lambda$ and the velocity $v_{A}$). Definite radiation mechanism for the solitons to power the FRBs remain to be identified. The total release of magnetic energy ultimately exciting the solitary wave is given in eq.(2), and even this may be an extreme limit. At $v_{A} \sim c$, the soliton would travel at several radii in $\leq 1 \, ms$ and could be the cause of the pulses.

\section{Gravitational wave from f-modes of the crust}

As suggested above, the very recent detection of FRB activity from the galactic magnetar SGR 1935+2154 (CHIME/FRB Collaboration 2020) opens the possibility of a future test of this class of models using gravitational waves. The scenario of magnetic instabilities cracking the crust suggests the excitation of crustal modes, akin to ripples in a pond. These are known as $f$-modes in the astroseismology works. On very general grounds (Thorne 1987, Ho et al. 2020) one can write down the amplitude of an oscillation decaying on a timescale $\tau_{gw}$ as $h(t) = h_{0} e^{-t/\tau_{gw}} \sin \omega_{gw} t$. Using the general definition of the gravitational wave luminosity $dE_{gw}/dt$ for an impulsive source at a distance $d$, an integration in time yields (Ho et al. 2020)

There are ongoing searches for gravitational wave sources from transient signals (Abbott et al. 2020, Klimenko et al. 2016) associated, for instance, with supernovae (Lundquist et al. 2019, Iess et al. 2020, Szczepanczyk et al. 2021, Skliris et al. 2020) and FRBs (. Recently, a gravitational wave transient candidate, S191110af, was detected, consisting of a signal at 1.78kHz and lasting for 0.104s (Chatterjee 2019). It was pointed that the frequency and duration of such an event are consistent with the fundamental stellar oscillation mode ($f$-mode, for short) of a NS with $sim 1.25M_\odot$ and $\sim 13.3$km (Kaplan et al. 2019). It can be seen, for instance, in Anderson and Comer (2001), van Eysden and Melatos (2008), Keer and Jones (2015), and Sidery et al. (2010) that $f-$mode oscillations can be triggered by transient events internal to the NS, such as magnetic field reorganization.

Note that the importance and interest that have been given to the $f-$mode associated gravitational waves lies in the fact that the mode frequency depends on the dynamical timescale of NSs, so that it is a probe of NS macroscopic features, as density, mass and radius (Anderson \& Comer 2001, Kokkotas et al. 2001, Anderson and Kokkotas 1996).

Considering a stellar oscillation with frequency $\nu_{gw}$ that damps on a time scale $\tau_{gw}$, the gravitational wave amplitude from such an oscillation is (Echeverria 1989, Finn 1992)

\begin{equation}
	h(t)=h_0e^{-t/\tau_{gw}}\sin(2\pi\nu_{gw}t).
\end{equation}
The peak luminosity $h_0$ can be determined from the integration of the gravitational wave luminosity of a source at distance $d$ (Owen 2010),

\begin{equation}
	\frac{dE_{gw}}{dt}=\frac{c^3d^2}{10G}(2\pi\nu_{gw}h_0e^{-t/\tau_{gw}})^2.
\end{equation}
Such an integration over $0<t<\infty$ naturally gives the total gravitational wave energy emitted, $E_{gw}$. By solving the integration for $h_0$ yields

\begin{eqnarray}
    h_{0} = 4.85 \times 10^{-18} {\biggl( {{10 kpc} \over {d}} \biggr)} {\biggl( {{E_{gw}} \over {M_{\odot} c^{2}}} \biggr)}^{1/2} \times \nonumber\\
    \times {\biggl( {{1 kHz} \over {f_{gw}}} \biggr)} {\biggl( {{0.1 \, s \over {\tau_{gw}}}} \biggr)}^{1/2} \,\,\label{} .
\end{eqnarray}

These gravitational wave modes must be excited by some mechanism, which in our case, as in (Sieniawska \& Bejger 2019, de Freitas Pacheco 1998), is the magnetoquake energy release.However, the excitation of the Alfven waves depends on the exact fracture
induced by the magnetic field, the state of the crust and other details. As a result, the energy effectively going into this propagation can be calculated in a detailed simulation only, and even so many
uncertainties would remain. Given that eq.(2) is an upper limit, if it is totally converted into
$f$-mode excitation, we obtain $h_0 \sim 3 \times 10^{-24}$ at $10kpc$ as an upper limit to the GW emission.

Studies of the $f$-modes have shown frequencies $f_{gw}$ between $1.25 - 2 kHz$ and a variety of damping times. However, it has been shown that the product $\omega_{gw} \tau_{gw}$ can be considered a function of the compactness $M/R$ only (Chirenti, de Souza \& Kastaun 2015). Therefore, a measurement of the frequency and damping time would be consistent with a variety of equations of state with a given fixed compactness. If the maximum released energy eq. (2) is effectively employed to excite the $f$-modes, the estimate $h_{0} \approx 3-4 \times 10^{-24}$ for the FRBs of SGR 1935+2154 at $10 kpc$, may be modified by observing that recent works (Zhou et al. 2020, Mereghetti et al. 2020) obtained a lower distance to the source. The design goals of the Advanced LIGO (LIGO Scientific Collaboration 2015) would be still insufficient, but the projected Einstein Telescope
could reach this figure (Huerta \& Gair 2010) although the detection of a closer source in a monitoring campaign and a careful treatment of the data around the time of the burst could improve the situation. On the other hand, $h_{0}$ would be within the reach of a project such as the {\it Einstein Telescope}, within the next decade (Maggiore et al. 2020). The detection of gravitational waves associated with FRBs would be important to support or discard magnetar models in its various versions, even good upper limits would be useful for this task.

\section{X-ray emission}

Although the previous evidence in favor of high-energy emission in coincidence with the FRBs was scarce, the detection of a burst of hard X-rays by the INTEGRAL mission (Mereghetti et al. 2020) from SGR 1935+2154 has added an important piece of evidence that must be addressed when discussing the generation of the FRBs by magnetars. The presence of X-rays slightly delayed (a few $ms$) from the FRB have been interpreted as evidence in favor of synchrotron maser emission, but there are other mechanism(s) which can also be invoked.

One could imagine that dissipation of the Alfven waves could be enough to power the observed X-ray spikes. However, a detailed study by Chen et al. (2020) suggests that, in spite that the plasma arranges itself to propagate beyond the charge-starved region, its dissipation falls short of power to explain the X-ray emission. Therefore, the X-rays should come form another process.
One of these proposals (Yuan et al. 2020) simulated the behavior of Alfven waves converted to plasmoids at high $z$ in the magnetosphere. The magnetic field lines, broken by the plasmoid ejection process reconnect behind the plasmoid ejection region and extract energy from the magnetosphere (i.e. a reservoir much bigger than the release in eq.(2)). As a general, quasi-dimensional estimate, the energy release by the reconnection process can be written as (see, for instance, Asai et al. 2002) $dE/dt |_{rec} = (B^{2}/{4 \pi}) v_{i} A$, where $v_{i}$ is the inflow velocity entering the reconnection zone and $A \sim L^{2}$ is the area of the reconnection happening up in the magnetosphere. The two well-studied mechanisms suggested for the physical picture of the reconnection, the ``slow'' Sweet-Parker (Sweet 1958, Parker 1957) and ``fast'' Petschek (1964) mechanisms have been found to be adequately described by $v_{i} \propto B^{1+\alpha}$, with $\alpha = 0.5$ and $\alpha = 1$ for the Sweet-Parker and Petschek proposals. The energy release estimate is

\begin{equation}
    {d E \over {dt}}|_{rec} = 2 \times 10^{39} {\biggl( {B \over {3 \times 10^{12} G}} \biggr)}^{2} {\biggl( {v_{i} \over {0.01 c}} \biggr)} {\biggl( {L \over {100 m}} \biggr)}^{2}erg \, s^{-1} \,\, ,
\end{equation}

mainly in X-rays and accelerated electrons. Detailed simulations (Werner \& Uzdensky 2017) suggest that $\sim 1/3$ of the energy contributes to accelerate particles and $\sim 2/3$ is dissipated in hard X-ray photons, with a spectrum which is not particularly important here. This figure must be compared with the inferred $1.4 \times 10^{39} erg$ obtained by Mereghetti et al. (2020) with INTEGRAL or the slightly higher value reported by Tavani et al. (2020) with AGILE, and shows that the propagation of the Alfven perturbation can trigger a suitable magnetospheric release, logically delayed from the FRB as observed.

\section{Discussion}\label{sec:dis}

We have considered in this work a scenario for the production of FRBs and related issues.
The recent confirmation of the galactic magnetar SGR 1935+2154 source at $\sim 10 kpc$ as the origin of
FRB pulses leads to believe that at least one type of source has been identified, and therefore
models of FRBs generation from them need to be re-examined. Starting with the idea that magnetic fields do not
achieve definitive configurations for $\sim$ centuries (Gourgouliatos \& Esposito 2018), we have considered their
dominant role in crust dynamics when their intensity is high enough (e.g. eq.(3)).
It is important to remark that the sources may not be restricted to the ``magnetar'' class. Empirical evidence for a link between young pulsars and FRBs has been presented by Nimmo et al. (2021), still requiring
high magnetic fields but also finding a range of timescales and luminosities. These authors then associate
the production of FRBs to young pulsars in general. For the magnetoquake scenario to work, however, it
would be enough that a {\it local} field reaches $B \sim 10^{15} \, G$, not necessarily the large-scale field
thought to be harboured by magnetars. In fact, and independently of our analysis, a patch $\sim 100 m$ size possessing higher multipole intensities in this ballpark is currently considered and found some support in direct observations using the NICER data (Bilous et al. 2019). A local cracking of the crust by this local
field would produce, in principle, FRBs with some variation in timescales and luminosities. This is why a closer look to the arrival directions of FRBs from known young pulsars would be potentially revealing.

The Alfven wave bunches are a simple, but not the only mechanism invoked to produce the FRBs. A discussion by Wang (2020)
addressed these possibilities and concluded that all them have some difficulty for their identification with the main
emission of FRBs. Specifically, the short duration predicted for the coherent curvature radiation. Within our simple picture we are not able to address this issue, which involves the detailed geometry of the field lines (simplified by our treatment) and
damping effects. Cooper and Wijers (2021), on the other hand, address positively the coherent curvature as a ``universal''
origin for a large range of luminosities, and suggest a scale of $\leq 10^{7} cm$, low inside the magnetosphere for its origination. On the other hand, Belobodorov (2021) argues for the ``choking'' of the FRBs if the emission originates at distances
$\ll 10^{10} cm$. The issue of locating the charge-starvation region, closer to the end of the magnetosphere becomes critical
in this sense, because it may or may not allow the observed radio emission at all.

Next we examined the propagation of a perturbation $\xi$ excited by plate ``blowout'' by the
magnetic field, by solving the simplest Green function in the magnetosphere $z > 0$. The
curvature radiation as the most likely mechanism for the production of the radio pulse was
shown to be consistent with the expected features started by the event. We have also pointed out
that ``spiky solitons'' (not treated in detail here) may be involved in the problem, although their detailed physics is even
more complicated and have not been clarified in this context. The detection of
diffuse radiation of a plerion-type (Resmi, Vink \& Ishwara-Chandra 2020) could also hold clues about the overall nature of the
radiation mechanism and its diffusion in the magnetospheric region, although not necessarily related
to the pulse events (in particular, the transition radiation through the propagating hump is found to be too
weak to contribute). The detection of a change in the spindown (Younes et al. 2020) after the FRBs in SGR 1935+2154 is another relevant clue, possibly interpreted a topology change of the magnetic field, which is quite
expected in our model. Reasons for a closely associated X-ray burst from the source (Mereghetti 2020) can be also accommodated (section VI), but it is premature to claim a convincing explanation. Finally, an estimation of
the gravitational signal expected from the crust $f$-modes was found to yield weak amplitudes, but likely
crucial to establish the reality of the whole picture once advanced facilities could be
operated. Meanwhile, the characterization of magnetars as sources of FRBs will continue, and
the discussion about the source(s) clarified by continuing observations and related
statistical considerations (Katz 2016).

\begin{acknowledgements}
PHRSM would like to thank CAPES for financial support. JEH has been supported by the CNPq Agency (Brazil) and the FAPESP foundation (S\~ao Paulo, Brazil).
\end{acknowledgements}

\appendix                  

\section{Review of the transmission of energy into the magnetosphere: basic equations of the physical picture}

The starting point of the calculations of the transmission of the released energy into the magnetosphere is the derivation of a suitable wave equation, by linearizing the motion and continuity equations about a static equilibrium. After a standard procedure, the result is (Blaes et al. 1989)

\begin{eqnarray}
	\rho\frac{\partial^2\xi}{\partial t^2}=\nabla .\delta\sigma+\frac{1}{c}\delta{\bf j}\times{\bf B}+\delta\rho{\bf g}-\nabla\delta p,\label{5}\\
	\delta\rho=-\nabla.(\rho\xi) \,\, .\label{6}
\end{eqnarray}

In eqs. (A.1) and (A.2), $\rho$ is the density profile, obtained from integrating the hydrostatic equilibrium equation, $\xi$ is the displacement of an element of material from its equilibrium position, $c$ is the speed of light, ${\bf j}$ is the current density, ${\bf g}=-g\hat{z}$ is the Newtonian gravitational acceleration and $p$ is the pressure due to degenerate electrons.

The components of the perturbed elastic stress tensor are

\begin{equation}\label{3}
	\delta\sigma_{ij}=\left(\kappa-\frac{2\mu}{3}\right)\delta_{ij}\nabla.\xi+\mu\left(\frac{\partial\xi_i}{\partial x_j}+\frac{\partial\xi_j}{\partial x_i}\right) \,\, ,
\end{equation}
in which $\kappa$ is the bulk modulus and the ions in the solid crust are arranged in a Coulomb lattice whose shear modulus is given by $\mu$. When writing eq.(A.3), terms which arise if there is a static elastic stress are neglected.

The crust is effectively a perfect electrical conductor of seismic frequencies. The perturbed electric field can be written in terms of $\xi$ as

\begin{equation}\label{4}
	\delta{\bf E}  -\frac{1}{c}\frac{\partial\xi}{\partial t}\times{\bf B} \,\, .
\end{equation}

Let us also write the perturbed Maxwell equations

\begin{eqnarray}	
	\nabla\times\delta{\bf E}=-\frac{1}{c}\frac{\partial\delta{\bf B}}{\partial t},\label{5}\\
	\nabla\times\delta{\bf B}=\frac{4\pi}{c}\delta{\bf j}+\frac{1}{c}\frac{\partial\delta{\bf E}}{\partial t} \,\, .\label{6}
\end{eqnarray}

In order to derive a simpler wave equation, Blaes et al. (1989)  have considered a vertically propagating shear wave polarized in such a way that $\xi \propto {\bf z}\times{\bf B}$. In this case, $\nabla.\xi=\delta\rho=\delta p=0$ and Eqs.(A.1-A.3), after standard vector calculus manipulations, now read

\begin{eqnarray}
	\rho\frac{\partial^2\xi}{\partial t^2}=\nabla .\delta\sigma+\frac{1}{c}\delta{\bf j}\times{\bf B},\label{7}\\
	\nabla\rho=0,\label{8}\\
	\delta\sigma_{ij}=\mu\left(\frac{\partial\xi_i}{\partial x_j}+\frac{\partial\xi_j}{\partial x_i}\right) \,\, .\label{9}
\end{eqnarray}

The resulting wave equation is

\begin{equation}\label{11}
	\frac{d}{dz}\left(\tilde{\mu}\frac{d\xi}{dz}\right)-\tilde{\rho}\frac{d^{2}\xi}{dt^{2}} = 0 \,\, ,
\end{equation}
with

\begin{eqnarray}
	\tilde{\mu}\equiv\mu+\frac{(B\cos\alpha)^2}{4\pi},\label{12}\\
	\tilde{\rho}\equiv\rho+\frac{B^2}{4\pi c^2} \,\, , \label{13}
\end{eqnarray}
being the effective shear modulus and density, respectively, and $\cos\alpha\equiv B_z/B$.

Up to now the wave equation applies to the crust and beyond. However, instead of studying the transmission across the crust (Blaes et al. 1989), we shall simply assume that a fraction of the total released energy eq.(2) makes its way into the magnetosphere being transmitted by the sudden motion at the base of the line, in the form of an impulsive perturbation. That is, we disturbed the field lines by assuming a sudden ``blowout'' of the local crust material patch with the field lines frozen in it in the form $f(z) \delta (t-t_{0})$ and seek for the response of the eq.(5) in Section 3. This kind of picture differs from the seismological approach in which the cracking of the crust is assumed. Here we rather envisage the magnetic instability lifting the base of the lines and seek to solve the Alfvenic wave propagation and its radiated power.

\label{lastpage}

\end{document}